\documentclass[conference, letterpaper]{IEEEtran} 
\IEEEoverridecommandlockouts
\usepackage{cite}
\usepackage{amsmath,amssymb,amsfonts}
\usepackage{algorithmic}
\usepackage{graphicx}
\usepackage{textcomp}
\usepackage{xcolor}
\usepackage{url}
\def\BibTeX{{\rm B\kern-.05em{\sc i\kern-.025em b}\kern-.08em
    T\kern-.1667em\lower.7ex\hbox{E}\kern-.125emX}}
\usepackage[linesnumbered, ruled]{algorithm2e} 
\usepackage{algorithmic}
\usepackage{xcolor}
\makeatletter
\makeatletter
\def\ps@IEEEtitlepagestyle{%
  \def\@oddfoot{\mycopyrightnotice}%
  \def\@oddhead{\hbox{}\@IEEEheaderstyle\leftmark\hfil\thepage}\relax
  \def\@evenhead{\@IEEEheaderstyle\thepage\hfil\leftmark\hbox{}}\relax
  \def\@evenfoot{}%
}
\def\mycopyrightnotice{%
  \begin{minipage}{\textwidth}
  \centering \scriptsize
  Copyright~\copyright~2024 IEEE. Personal use of this material is permitted. Permission from IEEE must be obtained for all other uses. The final version of record is published in the 9th International Conference on Big Data Analytics (ICBDA), 2024. The final version of this paper is available on IEEE Xplore. You can access it here: DOI: [10.1109/ICBDA61153.2024.10607167]. Available: \url{https://ieeexplore.ieee.org/abstract/document/10607167}
  \end{minipage}
}
\makeatother
\begin{document}

\title{DBNode: A Decentralized Storage System for Big Data Storage in Consortium Blockchains \\

\thanks{This work is funded by the German Federal Ministry of Education and Research (BMBF) under FT-Chain (01DS21011) project.}
}
\makeatletter
\newcommand{\linebreakand}{%
  \end{@IEEEauthorhalign}
  \hfill\mbox{}\par
  \mbox{}\hfill\begin{@IEEEauthorhalign}
}
\makeatother
\author{\IEEEauthorblockN{1\textsuperscript{st}  Narges Dadkhah}
\IEEEauthorblockA{\textit{Department of Mathematics and Computer Science} \\
\textit{Freie Universität Berlin}\\
Berlin, Germany \\
narges.dadkhah@fu-berlin.de}
~\\
\and
\IEEEauthorblockN{2\textsuperscript{nd}  Xuyang Ma}
\IEEEauthorblockA{\textit{School of Information and Communication Engineering} \\
\textit{University of Electronic Science and Technology of China}\\
Chengdu, China \\
xuyangm7@gmail.com}
~\\

\linebreakand

\IEEEauthorblockN{3\textsuperscript{rd} Katinka Wolter}
\IEEEauthorblockA{\textit{Department of Mathematics and Computer Science} \\
\textit{Freie Universität Berlin}\\
Berlin, Germany \\
katinka.wolter@fu-berlin.de}
~\\
\and
\IEEEauthorblockN{4\textsuperscript{th} Gerhard Wunder}
\IEEEauthorblockA{\textit{Department of Mathematics and Computer Science} \\
\textit{Freie Universität Berlin}\\
Berlin, Germany \\
g.wunder@fu-berlin.de}
}
\onecolumn 
\maketitle

\begin{abstract}
Storing big data directly on a blockchain poses a substantial burden due to the need to maintain a consistent ledger across all nodes. Numerous studies in decentralized storage systems have been conducted to tackle this particular challenge. Most state-of-the-art research concentrates on developing a general storage system that can accommodate diverse blockchain categories. However, it is essential to recognize the unique attributes of a consortium blockchain, such as data privacy and access control. Beyond ensuring high performance, these specific needs are often overlooked by general storage systems. This paper proposes a decentralized storage system for Hyperledger Fabric, which is a well-known consortium blockchain. First, we employ erasure coding to partition files, subsequently organizing these chunks into a hierarchical structure that fosters efficient and dependable data storage. Second, we design a two-layer hash-slots mechanism and a mirror strategy, enabling high data availability. Third, we design an access control mechanism based on a smart contract to regulate file access.
\end{abstract}

\begin{flushleft}
\begin{IEEEkeywords}
Blockchain, Hyperledger Fabric, Big Data, Erasure Coding, Access Control
\end{IEEEkeywords}
\end{flushleft}

\section{Introduction}

In recent years, extensive investigations have been conducted regarding using blockchain in various use cases to harness the technology's inherent characteristics of immutability, decentralization, and tamper-proof. Nevertheless, blockchain technology, which was introduced in 2008 with the launch of Bitcoin\cite{nakamoto2008bitcoin}, comes with some limitations. One of the most significant challenges is storage capacity. In certain use cases, there is a need to store large files and big data securely, and blockchain faces storage constraints.
Numerous solutions and research efforts have been undertaken to address this challenge. Each solution has its own set of pros and cons, such as excessive latency, lack of file access control, security and privacy challenges, and more. The common approach involves the use of off-chain distributed storage systems alongside the blockchain.
By contrast, this paper is focused on introducing a novel solution that resolves blockchain storage challenges within the blockchain nodes. This approach eliminates the need for external technologies and provides a comprehensive solution to overcome previous difficulties.

\subsection{Our Contribution}
Our research centers around two novel concepts: First, designing an innovative approach within the blockchain ecosystem that enables the storage of large files while maintaining a high level of security, ensuring data integrity, and tolerating node failures. 
Second, implementing a smart contract allows clients to define access rules for their files, control file availability, and set retrieval limits.
In order to implement these concepts, we utilized Hyperledger Fabric (HLF)\cite{10.1145/3190508.3190538}, a consortium blockchain, incorporated erasure coding technology, and introduced \textit{``DBNode''} a new type of node. Furthermore, to accomplish the objectives of this system, two-layer hash slots and mirror strategy in this architecture are designed, which will be explained in detail in this paper.
Additionally, a user-friendly HLF application has been developed to enable clients to store their large files as transactions within the blockchain. Clients are relieved of managing how their files are stored in the blockchain separately. They can easily transmit their files along with hyperparameters, and the blockchain takes care of the rest. 
In the evaluation section, we conducted a comparative analysis between DBNode and InterPlanetary File System (IPFS) \cite{DBLP:journals/corr/Benet14} as an off-chain storage solution in three different experiments. The results demonstrated that DBNode shows a significant improvement in file retrieval and effectively adapts to changes in network bandwidth.

\section{Related work}

The issue of storing big data within a blockchain continues to be an ongoing concern within this field. Several solutions have emerged to address this particular concern. One of these approaches is increasing the block size to handle larger files or data entries such as Bitcoin SV\footnote{https://bitcoinsv.com/}. Blockchain utilizes a replication system in which every node within the network stores all the blocks present in the ledger. As block size expands, the need for better network bandwidth arises to prevent propagation delay. Simultaneously, nodes require more resources to be able to store extensive data. 

Another issue of storing large amounts of data in some types of blockchains is a fee challenge. In cryptocurrencies like Bitcoin, these fees are calculated based on a fee rate per byte, which makes increasing block size for data storage impractical. Some research like \cite{wang2020blockchain} implements lightweight nodes that do not require storing a complete copy of the blockchain. This approach effectively reduces the storage overhead on these nodes. However, to access all the blocks in the ledger, these lightweight nodes need to connect to full nodes. One drawback of this method is that the load of the network is on the full nodes and can lead to increased communication costs within the network.

Earlier studies have investigated approaches to tackle this challenge by using well-known databases that handle large data effectively. Researchers integrated the advantageous attributes of these databases with the functionalities of blockchain technology in order to devise novel solutions.
BigchainDB\cite{mcconaghy2016bigchaindb} has introduced an alternative approach for addressing big data challenges. Rather than directly incorporating big data features into blockchain, they have utilized MongoDB as the foundational structure and integrated blockchain features such as decentralization and immutability into this framework. Mystiko \cite{8622341} introduces a private blockchain storage solution. The approach undertaken in this study is built upon the Apache Cassandra distributed database framework, and blockchain features are integrated into this storage system. To enhance scalability and minimize network overhead, this investigation employed sharding-based replication. Unlike full node replication, where data is duplicated across all nodes, the data selectively replicates among a specific number of nodes according to Cassandra's replication factor. Alternatively, a similar approach is explored in HBasechainDB\cite{sahoo2018hbasechaindb}, which is built upon Apache HBase and Hadoop.

Researchers in \cite{8726839} proposed an alternative solution to eliminate the need for full replication. They implemented an erasure code mechanism, dividing blocks into smaller fragments and distributing these fragments instead of relying on complete block replicas. Cloud servers are a common solution for storing large files but primarily rely on centralized authority. 

To mitigate the risks associated with centralization, such as single-point failure, researchers, as referenced in \cite{shahid2020blockchain, 9380312, kawaguchi2019application, naz2019secure} employ off-chain storage alternatives such as the IPFS. In this approach, the actual file remains external to the blockchain network to prevent unnecessary chain lengthening and reduce network load. Within the blockchain, only the hash of the file is stored, acting as a unique identifier that simplifies file retrieval from off-chain storage. The primary challenge in these types of approaches is the absence of access control over the data.

\section{Background}
\subsection{Consortium Blockchain}
Accessibility is the fundamental distinction between public and consortium blockchains. Public blockchains are open to everyone, whereas consortium blockchains with permissions restrict access. In this research, our primary focus is on Hyperledger Fabric, a consortium blockchain governed by the Linux Foundation. In HLF, organizations, also known as members, should be invited to access the network. However, it is essential to note that our research is not confined to this specific platform; it can be adapted to other types of blockchain.

In this study, we harness the advantages of HLF to develop our solution more effectively by leveraging its diverse components. Each component within HLF has a unique identity defined by a Certificate Authority (CA) in the form of a pair of public and private keys. Membership service providers (MSPs) allocate specific roles to these identities, making them recognizable within the network. This characteristic of the HLF enables us to store large files securely, monitor file usage, and gain deeper insights into the functioning of this architectural framework.

\subsection{Blockchain related Components}

\subsubsection{Nodes}
\label{Nodes}

Within HLF, distinct components play crucial roles, such as the channel, organization, and three essential types of nodes:
\begin{itemize}
\item Peers: Peers play a fundamental role in the HLF framework. These nodes receive transactions within a block and verify their validity. If the transactions are validated, peers store them in their individual databases and subsequently update the ledger. Furthermore, peers can take on the role of endorsers. In this scenario, when a client submits a transaction to the blockchain, endorsement peers evaluate the transactions and ensure the authorization of the client within the network. Following this evaluation, they execute the transaction and then transmit the signed result to the client.
\item Client: Clients are initiators of transactions that are intended to be stored in the blockchain. The client sends a transaction proposal to the endorsement peers. Following the network's endorsement configuration file, once the client receives satisfactory responses from endorsement peers, it proceeds to dispatch the transaction to the ordered node.

\item Orderers: Orderers are responsible for generating the block with transactions and subsequently distributing the block across the peers. Given the execute-order-validate consensus mechanism in HLF, this node plays a vital role in upholding consensus throughout the network.
\end{itemize}

\subsubsection{Channels and Chain Codes}
\label{Channels and Chain Codes}
In HLF, a channel represents a subset of the ledger, and only members within that channel have access to the data contained within it. This feature proves valuable in situations where there are competitors in the network or transactions should be access limited to a specific group of network members. Within each channel, it is possible to develop different smart contracts. Smart Contract (SC) is business logic and a set of logical rules that execute automatically. In HLF, chaincode manages the packaging and deployment of these smart contracts.
\subsection{Erasure Coding}
Erasure coding is a common technology used for data redundancy in the field of storage. The most widely adopted erasure code is the Reed-Solomon (RS) code~\cite{reed1960polynomial}. Most popular storage systems such as Google File System~\cite{ghemawat2003google}, Windows Azure Storage~\cite{huang2012erasure} and Hadoop File System~\cite{shvachko2010hadoop} adopt this technology to guarantee the availability of the data. To store a file, this method divides a file into small chunks. As illustrated in Fig.~\ref{platform_diagram}, a (\(n, k\)) RS code encodes a piece of data that is composed of $k$ chunks into $n$ chunks, $k$ of which are original data chunks and $n-k$ of which are coded parity chunks. These $n$ chunks are organized as a ``stripe'' and then distributed to $n$ different nodes for storage.

In contrast to a full replication system, this method enables the recovery of the original data using any $k$ out of the $n$ chunks. Hence, a (\(n, k\)) RS code is able to tolerate \(n-k\) node failures while incurring a redundancy of \(\frac{n}{k}\times\). As a comparison, the traditional replication technology incurs a redundancy of \((n-k+1)\times\) to tolerate the same number of node failures.

\label{Erasure Coding}
 \begin{figure}[!htbp]
   \centering
    \includegraphics[width=9cm, height=4cm]{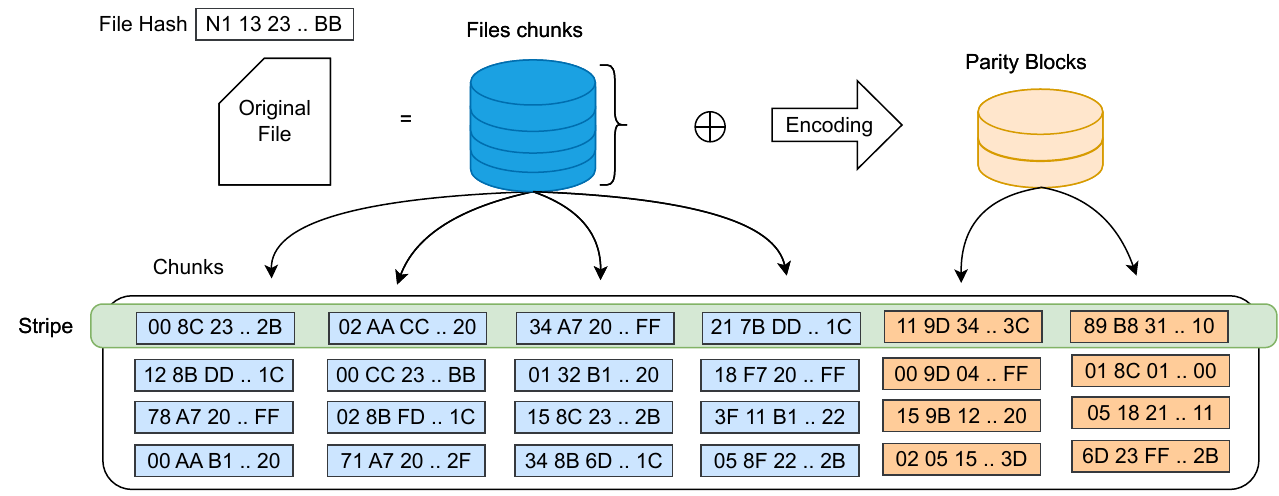}
     \caption { Erasure Coding}
     \label{platform_diagram}
\end{figure}

\section{Motivation}
\subsection{Distributed Storage Systems}
\label{ipfs}

Prior to explaining the DBNode structure, it is imperative to highlight some concerns regarding the use of off-chain storage systems in blockchain that motivated our research. IPFS is a well-known distributed storage system that has been used in several research such as \cite{passerat2019blockchain}, to address the storage limitations of blockchain.
IPFS, as a peer-to-peer decentralized file storage system, operates by breaking files into smaller pieces and generating unique hash values for each block. These hash values are then interlinked to be stored in a Merkle DAG, with a unique cryptographic hash for the entire file serving as the File ID (CID). In this network configuration, accessibility is open to all, facilitating participation in storing these file blocks. The CID, functioning as a persistent and unique identifier, simplifies the retrieval of the original file from the external off-chain storage system. Clients seeking to store their data on the blockchain create a transaction that includes this CID and then send this transaction to the blockchain. 

However, while off-chain mechanisms like IPFS offer solutions for addressing blockchain storage limitations, they introduce specific trade-offs and drawbacks:
\begin{itemize}
  \item Reduction in Security: A significant concern in combining blockchain with these mechanisms, particularly in systems like IPFS, is the absence of robust access control. Access control refers to controlling and restricting who can access or modify stored content. Unlike consortium blockchain, which offers fine-grained access control mechanisms, IPFS does not inherently offer robust security features. Detecting adversarial nodes and preventing their access within the IPFS network is one of the challenges in this mechanism. Blockchain has no control over how files are shared among IPFS nodes, which can result in security vulnerabilities.
  \item Privacy Concerns: Privacy is another area of concern. Before storing transactions in a blockchain, clients must first gain the CID from IPFS and then send the hash of the file to the blockchain. During this transmission process, if someone gains access to the hash of the file, they could potentially access the actual file. While encryption and access control measures can be implemented, the default configuration may not offer comprehensive privacy safeguards.
  \item Complexity: Setting up and managing IPFS, especially for users who are not familiar with decentralized systems, can be complex. Users typically need to employ a multi-step process, including uploading files to IPFS separately and then linking the resulting hash to the blockchain. This process may introduce redundancy and security risks and might not be as user-friendly for those primarily focused on interacting with the blockchain.
   \item Reading Latency: Another challenge with IPFS is reading latency, which has been evaluated in several research studies to determine the efficacy of file retrieval using IPFS. According to \cite{abdullah2021performance}, IPFS shows significant latency when reading large files compared to other technologies, such as File Transfer Protocol (FTP).
\end{itemize}  
In this investigation, we introduce a system compared with the IPFS solution, in which blockchain transactions contain the hash of the file while the actual file is stored within the blockchain. This innovative approach aims to simplify the management of large files and encapsulate the complexities of external storage mechanisms from clients.  Furthermore, our system seamlessly integrates the efficiency of IPFS for file storage and retrieval with the inherent attributes of blockchain, including immutability and traceability, resulting in a resilient and adaptable solution. 
\section{System Architecture}
\subsection{System Overview }
Before going into a comprehensive description of our architecture, we would like to provide an overview of the entire system. This paper introduces a novel node type called \textit{``DBNode''} within the HLF framework to efficiently address the challenge of storing large files. 
In HLF, applications interact with a blockchain in multiple ways, such as submitting transactions. Within the scope of this research, we developed an application based on erasure coding mechanisms, and we have established a dedicated channel known as File Channel (FC). Each organization participating in this network initially deploys one or more nodes as DBNodes within this storage system. Among the DBNodes in each organization, the one with better network conditions will be selected as the master node and registered in the FC channel. 

Subsequently, clients utilize our application to transmit their large files and specific hyperparameters that serve as access rules. Before client transactions are stored within the HLF, their files are divided into distinct chunks using erasure coding technology. Each chunk is assigned a unique identifier and organized into different fixed-size stripes. These unique identifiers adhere to hierarchical tree-like structure rules and create a unique hash ID in the header of this tree for the file (FID). Our application stores this file tree and distributes these stripes among the DBNodes within the FC channel.

In the next step, we have designed a smart contract within this channel to fulfill two primary responsibilities. Firstly, it creates a table used as a map to indicate which parts of the stripes will be stored in which organization. Secondly, this smart contract governs file accessibility while enforcing predefined rules. Also, within each organization, the master node will be responsible for maintaining a table showing which chunks of the stripes will be stored in the inner organization's DBNodes and distributing these chunks accordingly.

Finally, each DBNode stores the chunks in their respective databases. Once all the chunks are successfully stored within the DBNodes, the application transmits the file tree along with the client's hyperparameters as a transaction to the smart contract to store information in the HLF. This strategy marks the beginning of subsequent steps for the permanent storage and retrieval of files, as governed by the smart contract.

\hfill
\label{db}
 \begin{figure}[!htbp]
   \centering
     \includegraphics[width=11 cm, height=7 cm]{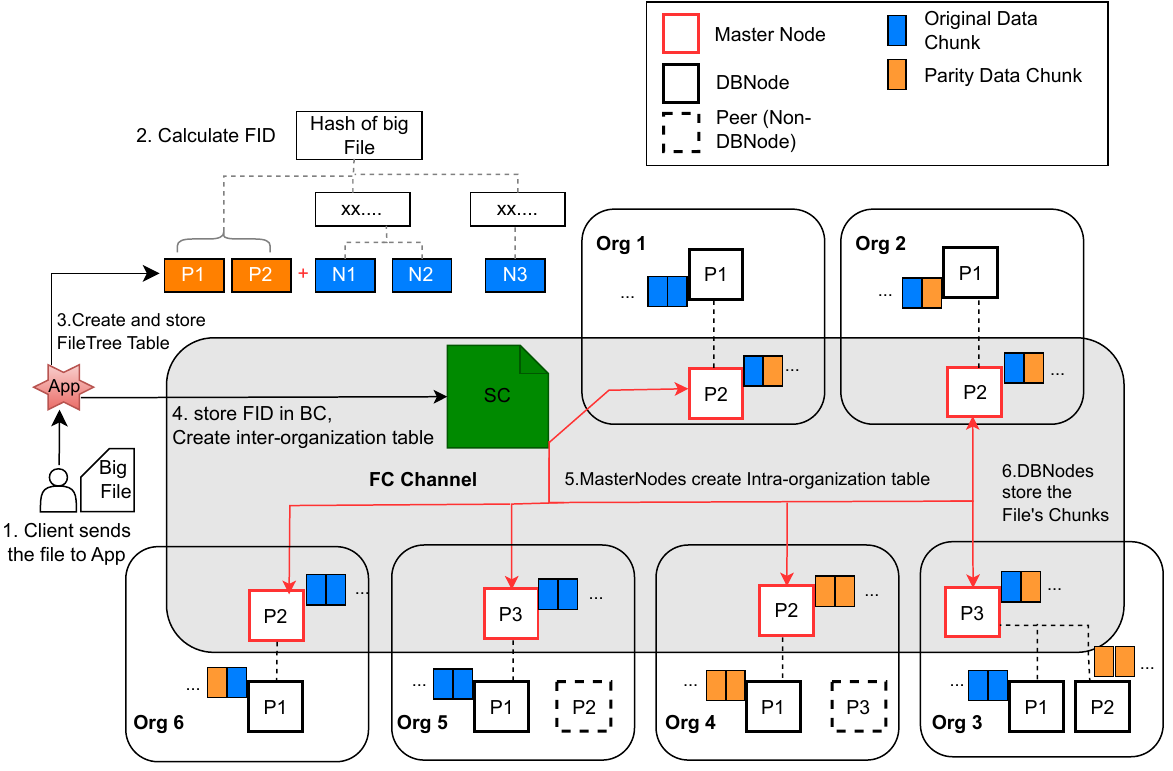}
     \caption { DBNode Architecture}
     \label{db-Architecture}
\end{figure}

\subsection{Erasure Coding Design}
\label{subsec: erasure coding}
Rather than simply using RS code directly, we consider specific characteristics of the consortium blockchain and design a suitable erasure coding. A consortium blockchain is composed of different organizations. In practice, an organization is usually associated with a company. Except for tolerating a certain number of node failures, the proposed storage system must also tolerate a certain number of organization failures as well.

Suppose a consortium blockchain consists of $N$ DBNodes and $M$ organizations. Each organization consists of \(\frac{N}{M}\) DBNodes. The maximum number of node failures that can be tolerated is $x$. The maximum number of organization failures that can be tolerated is $y$. In order to tolerate organization failures, $n$ encoded chunks within each data stripe will be divided into $l$ groups. Chunks belonging to different groups must be stored in different organizations. Then, we can derive the following set of constraints for setting a (\(n, k\)) erasure code:

\begin{equation}
    \label{cnstr1}
    n \leq N, l \leq M, \Big\lceil \frac{n}{l} \Big \rceil \leq \frac{N}{M}
\end{equation}
\begin{equation}
    \label{cnstr2}
    n-k \geq x
\end{equation}
\begin{equation}
    \label{cnstr3}
   \Big\lceil \frac{n}{l} \Big\rceil \cdot y \leq n-k
\end{equation}

Eq.~\eqref{cnstr1} specifies the constraints concerning the number of nodes.  A (\(n, k\)) erasure code can tolerate up to  \(n-k\) node failures. Hence, as shown in Eq.~\eqref{cnstr2}, this number must be equal to or greater than the maximum number of node failures. Each organization is responsible for storing up to \(\lceil \frac{n}{l} \rceil\) chunks. Therefore, the total number of failures must not exceed \(n-k\), as demonstrated in Eq.~\eqref{cnstr3}. Any erasure code that satisfies these three aforementioned constraints can be chosen. In reality, the members of the consortium blockchain should collectively decide which erasure code to employ. In this research experiment, we set up three organizations, each containing two DBNodes. Our assumption is that the system must tolerate three node failures and one organization failure, a configuration commonly encountered in other storage systems such as ~\cite{afike2010, ovsiannikov2013quantcast, shvachko2010hadoop}. Therefore, we have used the (6, 3) RS code for this purpose.



\subsection{Two-layer Hash Slots}
\label{subsec: slots}
Considering the characteristics of the network topology of a consortium blockchain, the proposed storage system consists of two layers: the inter-organization layer and the intra-organization layer. In each organization, a DBNode with the best network condition (e.g. highest bandwidth) will be selected as the master node of this organization, such as P3 illustrated in Fig.~\ref{hash-slots}. These master nodes are responsible for communication between organizations, thereby forming the inter-organization layer. Inside an organization, the master node needs to communicate with other common DBNodes, thereby forming the intra-organization layer.


\begin{figure}[!htbp]
\centering
     \includegraphics[width=11cm, height=6 cm]{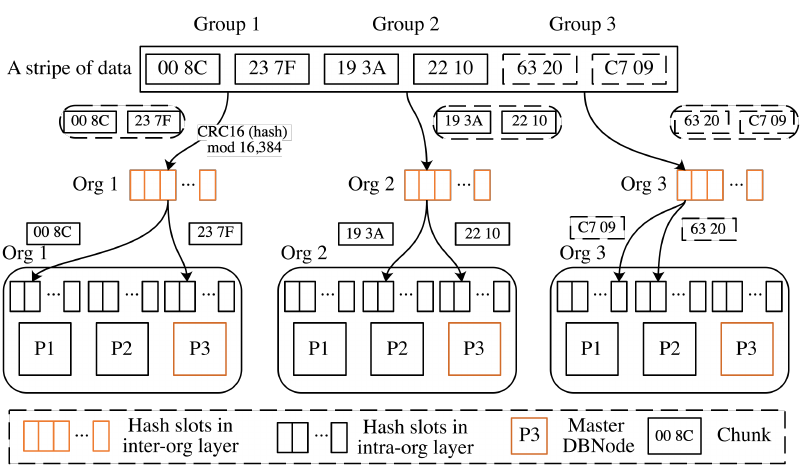}
     \caption {Two-layer hash slots}
     \label{hash-slots}
\end{figure}


As mentioned in Sec.~\ref{subsec: erasure coding}, in practice, an organization is typically owned by a single company, and the servers or nodes within the same organization are often located in the same data center. Consequently, network communication within an organization is usually significantly faster than communication between different organizations. To determine the appropriate node for storing each chunk, we have designed two hash slot tables. The hash slot is a widely used technology in distributed databases for distributing data uniformly. Database operators will allocate different numbers of slots to different storage nodes. Once a client requests to store a file, it can use the file hash to calculate which slot this file belongs to, and then this file will be stored in the corresponding storage node in the organization.

As illustrated in Fig.~\ref{hash-slots}, there are 16,384 hash slots in total, which is the same as the setting in Redis~\cite{carlson2013redis}. The first hash slot table is built for the inter-organization layer. According to the bandwidth of each master node (e.g. P3), different numbers of slots are allocated to different master nodes. The second hash slot table is built for the intra-organization layer. According to the storage capacity of each DBNode inside an organization, different numbers of slots are allocated to different DBNodes. Note that the allocation of hash slots follows different strategies in the two tables. In the inter-organization layer, the communication quality is more significant, and the allocation is determined based on the measure of bandwidth across various master nodes. In the intra-organization layer, network communication is much faster. Hash slots are allocated according to the storage capacity of each node to guarantee the balance of storage.


\subsection{Mirror Strategy}
\label{mirrorstrategy}

As explained in Sec.~\ref{subsec: erasure coding}, we select the parameters of the erasure code to tolerate node and organization failures. This necessitates distributing different data chunks across various nodes to achieve redundancy, which can sometimes conflict with the concept of hash slots. Fig.~\ref{hash-slots} illustrates the most ideal scenario where each chunk is stored in distinct DBNodes. However, based on the calculation result, there is a possibility that two chunks should be stored in the same node. For instance, after calculation, the first two chunks may be allocated to the same hash slot. Such outcomes from hash slot calculations can potentially violate the data redundancy requirement. In order to tolerate node failures and follow the rule of hash slots simultaneously, we design a ``mirror'' strategy to address conflicts, as shown in Fig.~\ref{mirror}. 

\begin{figure}[!htbp]
\centering
     \includegraphics[width=14 cm, height=4 cm]{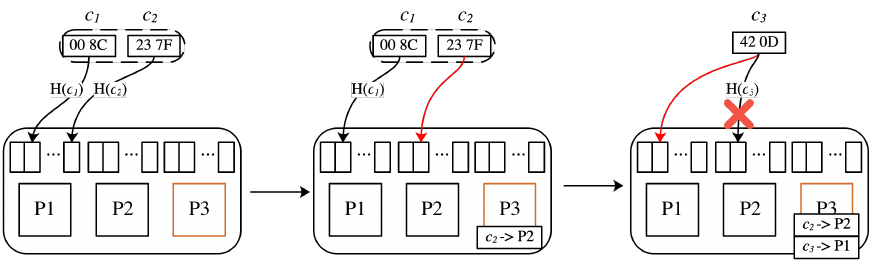}
     \caption {The process of mirror strategy}
     \label{mirror}
\end{figure}

According to the hashing results, both chunk $c_1$ and chunk $c_2$ should be stored in the same DBNode P1. The master node P3 notices this conflict and then stores chunk $c_2$ in a different node, i.e. P2. Simultaneously, a link $c_2 \rightarrow \textsc{P2}$ is stored in P3. When a client requests $c_2$, it will send the request to P1 according to the calculation result. Then, P1 will respond with the actual storage node P2 after searching links. In the future, if a chunk $c_3$ should be stored in P2, according to the mirror strategy, $c_3$ will be stored in P1, and a link \(c_3 \rightarrow \textsc{P1}\) will be stored in the master node P3. At this point, the mirror operation ends.

The mirror strategy addresses the conflict between data redundancy and hash slots by introducing links. Additionally, this strategy allows the client to specify the node or organization where a file should be stored, thus preserving data privacy. For instance, if a file must remain confidential within a specific organization, the client can specify that this organization is not permitted to store the file. In such cases, DBNodes within that organization will only store links instead of the actual data, ensuring that curious DBNodes in that organization cannot reach the data. Furthermore, an access control mechanism, which will be introduced in Sec.~\ref{Access Rule}, will specify which clients are granted access to the file.

By combining the mirror strategy with access control, neither curious DBNodes nor unauthorized clients can view private data.
The cost associated with the mirror strategy involves storing these links. However, each link is composed of a hash of a chunk and a public key of the DBNode. The size of a link is significantly smaller than the size of a chunk. Hence, this cost is considered acceptable. Such cost is evaluated in our experiment.


\subsection{Access Control}
\label{Access Rule}
In this research, we implement access rules using smart contracts to be able to design file accessibility. When clients intend to store a file, they can establish specific access rules within the blockchain transaction as hyperparameters. These parameters allow for fine-grained control over file access. When clients contribute to the blockchain to store the file alongside the FID, they have the ability to define access rules based on these component identities and their roles. These rules determine which nodes are granted access and which are restricted from accessing the file.

Specifically, three kinds of access rules are defined in this smart contract.
\begin{enumerate}
    \item Permission Rule: When storing a file, the client can set a permission list bound to this file. The permission list contains all the identities that are allowed to access this file.
    \item Banned Rule: In contrast, the client can also set a banned list for the stored file. The banned list contains all the identities that are not allowed to access this file.
    \item Token Rule: The client can assign any number of tokens to the stored file. Once someone requests this file, the number of tokens associated with this file will be reduced by one. If all the tokens have been used, this file and related metadata, e.g. file hash, will be removed by the smart contract and DBNodes, and no one can reaccess this file. 
\end{enumerate}



\label{Algorithm-Write}
\subsection{Implementation}
This section provides a more detailed explanation of algorithms employed for the purpose of reading and writing files in this research. The organization $M$ comprises $N$ nodes and $\mathcal{N}_k$ possesses its own database. As part of the architecture, at least one of these nodes $\mathcal{N}_i$ is introduced as a storage node within the network.
\begin{algorithm}[!htbp]
\begin{algorithmic}[1]
\STATE \textbf{Input:} Client sends a file with access rules. \\
\textbf{Parameters:} node ID $\eta$, file $F$, access rules $AC$, RS code ($n$, $k$).
\STATE \textbf{Initialize:} Divide the file $F$ into $J$ stripes and $\mathcal{C}_j$ chunks.
\FOR{each stripe \(S_i\), \(i=1,2, \cdots, J\)}
\STATE Use the RS code to encode \(S_i\) into $n$ chunks.
\STATE Get the hash slot table $HST$ from the smart contract.
\FOR{each chunk \(C_j\), $j = 1, 2, ..., n$}
\STATE Calculate the DBNode ID $\gamma$ that needs to store
 $C_j$ according to the chunk hash $H(C_j)$ and $HST$.
 \STATE Send the chunk $C_j$ to the DBNode \(\gamma\).
\ENDFOR
   \IF{storing is successful}
     \STATE Create a file tree $\mathit{T}_f$ that includes the hashes of the file, stripes and chunks.
     \STATE Upload the $\mathit{T}_f$ to the blockchain.
    \ENDIF
\ENDFOR
\RETURN The file hash $H(F)$
\end{algorithmic}
\caption{DBNodes Write File}
\label{alg:write}
\end{algorithm}

\label{Algorithm-Read}

\begin{algorithm}[!htbp]
\begin{algorithmic}[1]
\STATE \textbf{Input:} Client $C$ requests the file $F$ by sending the file hash $H(F)$.

\textbf{Parameters:} the file hash $H(F)$.
\STATE Fetch the file tree $T_f$ from the blockchain, and obtain a list of stripeHashes $S_f$.
\STATE Get the hash slot table $HST$ from the smart contract.
\FOR{each stripe hash $H(S_i)$ in $T_f$, $i = 1, 2, \cdots, J$} 
\FOR{each chunk hash $H(C_j)$ belonging to $S_i$, $j=1, 2, \cdots, n$}
 \STATE Calculate the DBNode ID $\gamma$ that stores $C_j$ according to $H(C_j)$ and $HST$.
 \STATE The client sends a request for $C_j$ to the DBNode $\gamma$.
\ENDFOR
\IF{the client has received $k$ chunks}
\STATE Decode these chunks to recover the stripe $S_i$. 
\ENDIF
\ENDFOR
\STATE Combine $J$ stripes to recover the file $F$.
\RETURN the file $F$
\end{algorithmic}
\caption{DBNodes Read File}
\label{alg:read}
\end{algorithm}

\section{Evaluation}
We use Hyperledger Fabric to deploy a consortium blockchain consisting of four organizations on a Macbook Pro with 32 GB memory and an Intel i5 CPU. Each organization contains three DBNodes. We implement DBNode in Golang and use gRPC to enable communication between different nodes. Each DBNode is running as a container. We use Pumba~\cite{pumba2023}, which is a chaos testing tool for docker, to emulate realistic network conditions. We designed three experiments to evaluate the proposed system. 

In the first experiment, we used DBNode to store different numbers of chunks from 0 to 1,000. 
This experiment evaluates the storage overhead of the proposed mirror strategy by examining the maximum number of links stored in a single DBNode and the total number of links stored across the entire system.

\begin{table*}[h]
\caption{Different sizes of files and corresponding real models.\label{file_size}}
\centering
\begin{tabular}{|c|c|c|c|c|c|c|}
\hline
File Size (MB) & 10          & 20         & 30          & 100  & 200    & 300   \\ \hline
Model     & MobileNetV3 & SqueezeNet & DenseNet121 & BERT & YOLOv3 & GPT-2 \\ \hline
\end{tabular}
\end{table*}
In the second experiment, we conduct a comparison of write and read latencies between the IPFS and DBNode systems. For this purpose, we allocate the same 1,000 Mbps bandwidth to each DBNode and write and read files with different sizes. Considering the machine learning model sizes in the real world, the files with different sizes are illustrated in Table~\ref{file_size}. The files from 10 MB to 30 MB represent small models that are commonly used in lightweight devices, e.g. mobile phones. The files from 100 MB to 300 MB represent large models that are commonly deployed on computers or servers.

In the third experiment, we allocate varying bandwidths to DBNodes in different organizations. We employ a stepped bandwidth setting by limiting the bandwidth of DBNodes in four organizations to 400, 800, 1,200, and 1,600 Mbps, respectively. The average bandwidth remains the same as in the second experiment.  We continue to use both IPFS and DBNode to write, store and test latencies.

For each experiment, 20 trials are conducted, and as the outcome, the mean value is reported. In each trial, we generate a file by filling it with random characters, and the nodes responsible for writing or reading the file are randomly selected. In the rest of this paper, the results of each experiment will be discussed in more detail.


\subsection{The Storage Overhead of Mirror Strategy}
\begin{figure}[!htbp]
\centering
     \includegraphics[width=10 cm, height=6 cm]{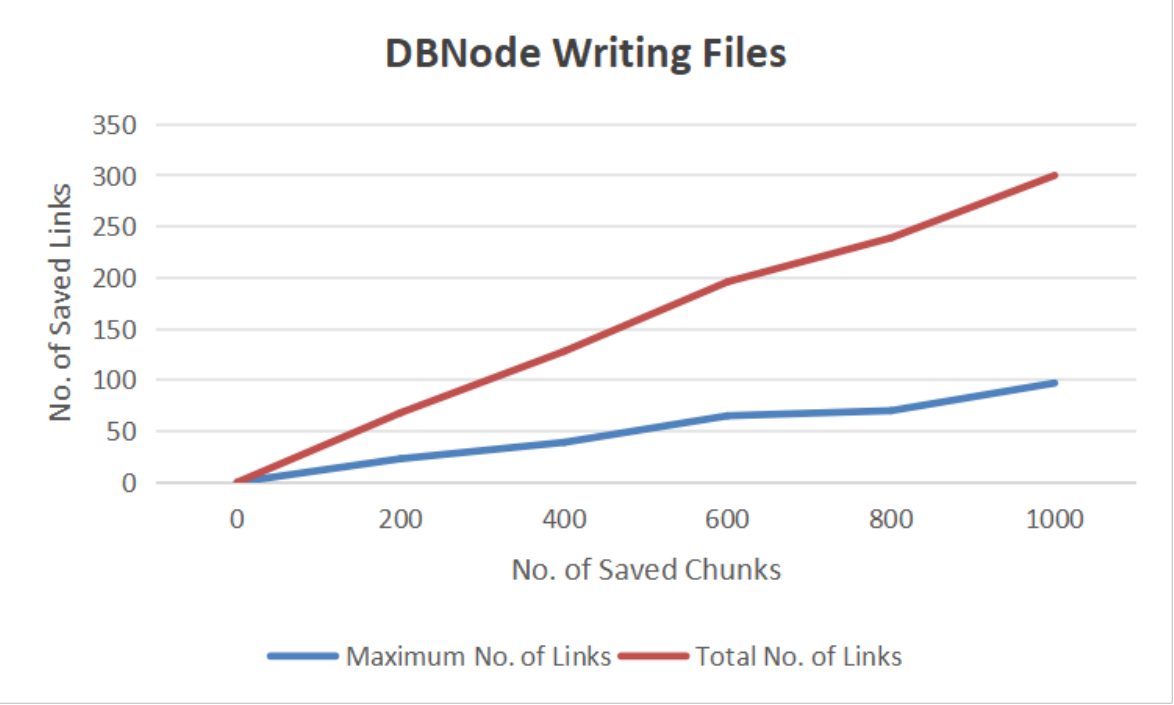}
     \caption {The maximum and total number of links that are stored in DBNodes}
     \label{link}
\end{figure}

As described in Sec.~\ref{mirrorstrategy}, the mirror strategy plays an essential role in the DBNode system. It helps resolve the conflict between the hash-slot rule and the data redundancy requirement. Moreover, this strategy preserves data privacy combined with the smart contract for access control.
However, the mirror strategy provides these benefits at the cost of additional storage requirements by introducing links.

The claim made in Sec.~\ref{mirrorstrategy} regarding the acceptability of the cost of the mirror strategy needs to be demonstrated in this experiment. Hence, we store different numbers of chunks in DBNodes to determine how much storage is required to store links. Two metrics, i.e. the maximum number of links stored in a single DBNode and the total number of links stored in the system, are concerned. The former shows the upper bound on the storage overhead of a single node. The latter represents the additional storage overhead required for the entire system.

The result of this experiment is illustrated in Fig.~\ref{link}. The number of links required to be stored grows linearly with the number of saved chunks. When storing 1,000 chunks in DBNodes, a single node stores about 100 links, and the whole system stores about 300 links. In this experiment, we use (6,3) RS code, and the size of each chunk is 1MB. A link is composed of a 64-byte hash value and a 64-byte public key. Therefore, a DBNode requires at most 12,800 bytes of disk space, and the whole system requires 38,400 bytes of disk space to store links while storing a 500 MB file. For each file stored, a single DBNode needs to spend up to \text{0.02\textperthousand} of the file size to store links while the entire system needs to spend approximately \text{0.06\textperthousand} of the file size. Such storage overhead is negligible compared to the file size.

\subsection{Writing and Reading Latency under the Same Bandwidth Setting}
\begin{figure}[!htbp]
    \centering
     \includegraphics[width=10 cm, height=6 cm]{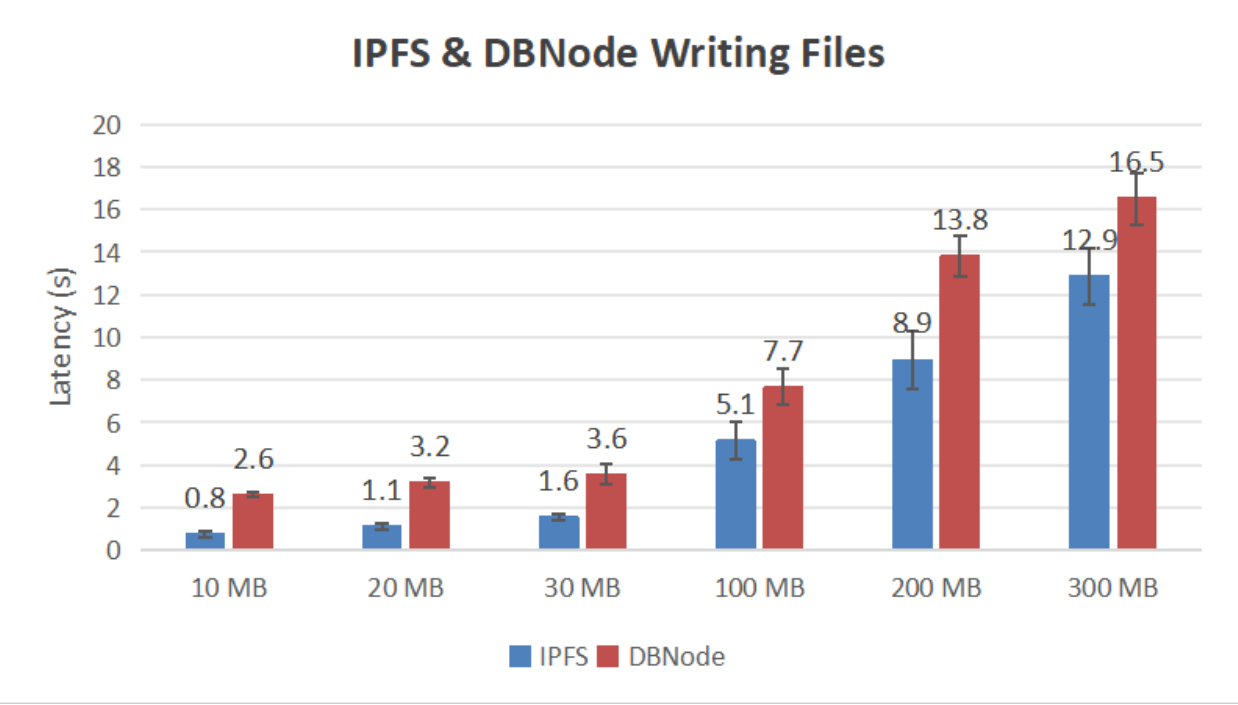}
     \caption {Writing latency of IPFS and DBNode}
     \label{sbw_w}
\end{figure}
\begin{figure}[!htbp]
    \centering
     \includegraphics[width=10 cm, height=6 cm]{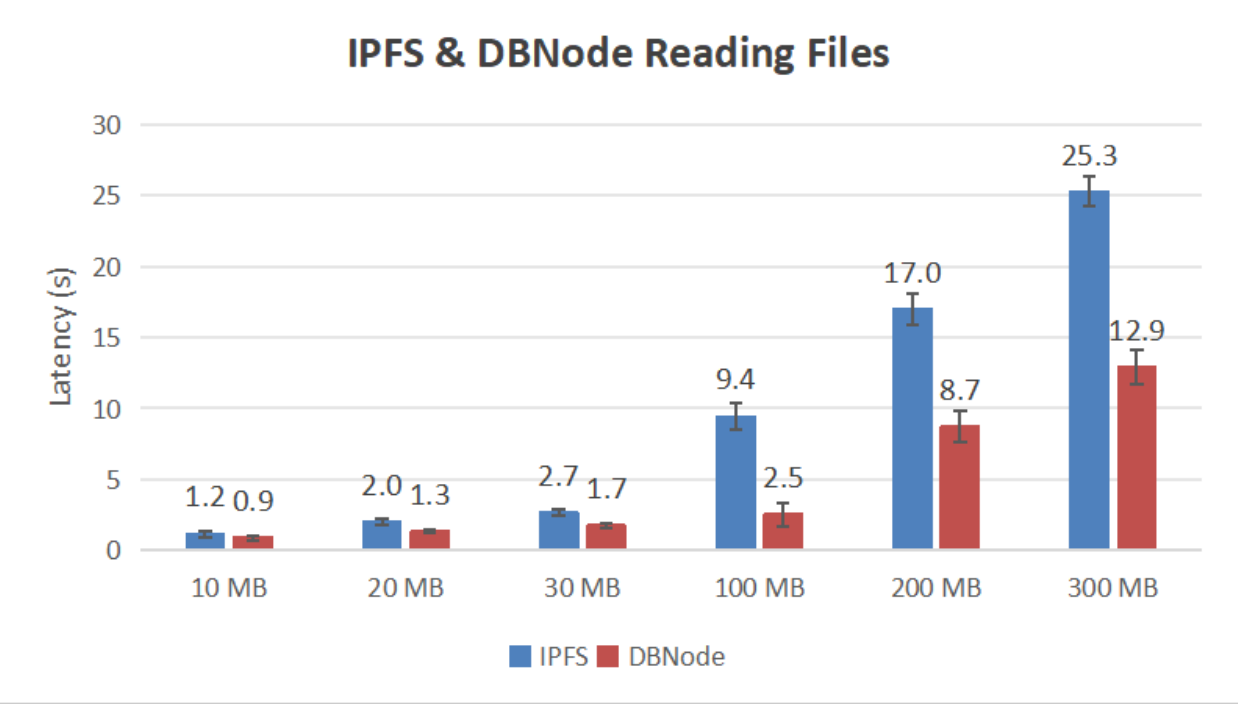}
     \caption {Reading latency of IPFS and DBNode}
     \label{sbw_r}
\end{figure}


In this experiment, we allocate equal bandwidth to each node and compare the write and read latencies between the DBNode system and IPFS. This comparison is essential as our system is designed to overcome the shortcomings of IPFS.

Fig.~\ref{sbw_w} illustrates the writing latency of the two systems. It can be observed that the DBNode system takes more time to store the file compared to IPFS. IPFS nodes will not actively back up data. They only transfer data in response to requests from other nodes. The node will keep one copy of the file after it gets this file. Therefore, the writing operation is performed locally, which is very fast. However, if an IPFS node fails shortly after adding a file to the IPFS network, this file will be unavailable to other nodes. 
The availability of a file in IPFS highly depends on how many nodes have requested it. Instead, the DBNode system leverages erasure coding to partition the file and stores each chunk in a different DBNode to obtain high availability. The availability of files in DBNodes is fixed based on the setting of erasure coding. Moreover, the file tree, which records the hash value of each chunk, is uploaded to the smart contract to guarantee that every chunk is searchable. The communication between the client and blockchain and the client and DBNode takes additional time, leading to increased writing latency. It is important to acknowledge that the process of storing files in the network occurs once for each file. Consequently, the minimal latency observed can be disregarded in contrast to achieving the assurance of file availability and accessibility, along with implementing access control mechanisms for the file.

Fig.~\ref{sbw_r} illustrates the reading latency of these two systems. The reading latency of the DBNode system is much less than IPFS. Such an advantage is even more obvious when reading large files (100, 200, 300 MB). Two primary reasons lead to this result. First, IPFS adopts the distributed hash table technology. When requesting a file, the node needs to take additional time to search the addresses of other nodes. A DBNode gets the hash slot table from the blockchain and calculates the address of another node locally. Since the hash slot table is already stored in the smart contract, the DBNode can find the table from the blockchain ledger of a peer. This process is speedy because the DBNode and the peer belong to the same organization or even run in the same physical machine. 

Second, the DBNode system adopts a ($n$, $k$) erasure code to partition a file into $n$ chunks. When requesting the file, the client can leverage multi-threading technology to request chunks from multiple nodes simultaneously. The file can be recovered once $k$ of $n$ chunks are obtained. However, in IPFS, the number of file copies depends on how many nodes request it. Especially for files that are not backed up by many nodes, IPFS nodes cannot take advantage of multi-threading to obtain data.  

\subsection{Writing and Reading Latency under the Stepped Bandwidth Setting}
\begin{figure}[!htbp]
    \centering
     \includegraphics[width=10 cm, height=6 cm]{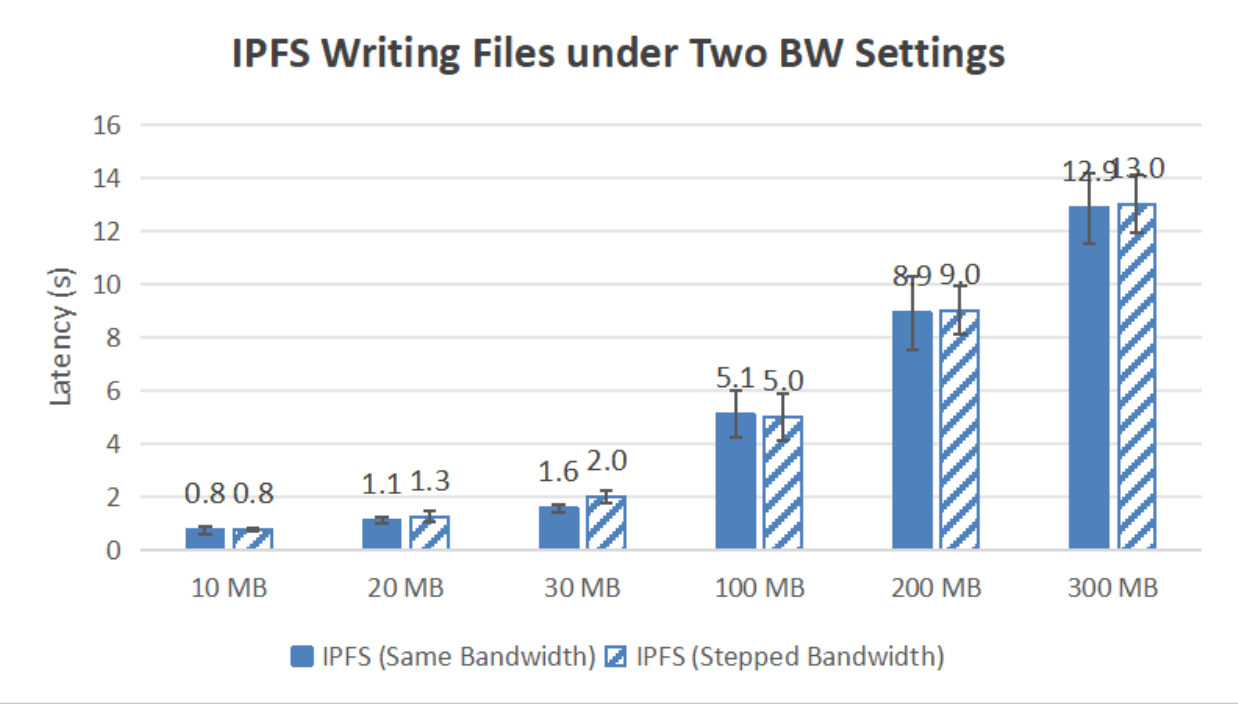}
     \caption {Writing latency of IPFS under different bandwidth settings}
     \label{ipfs_w}
\end{figure}
\begin{figure}[!htbp]
    \centering
     \includegraphics[width=10 cm, height=6 cm]{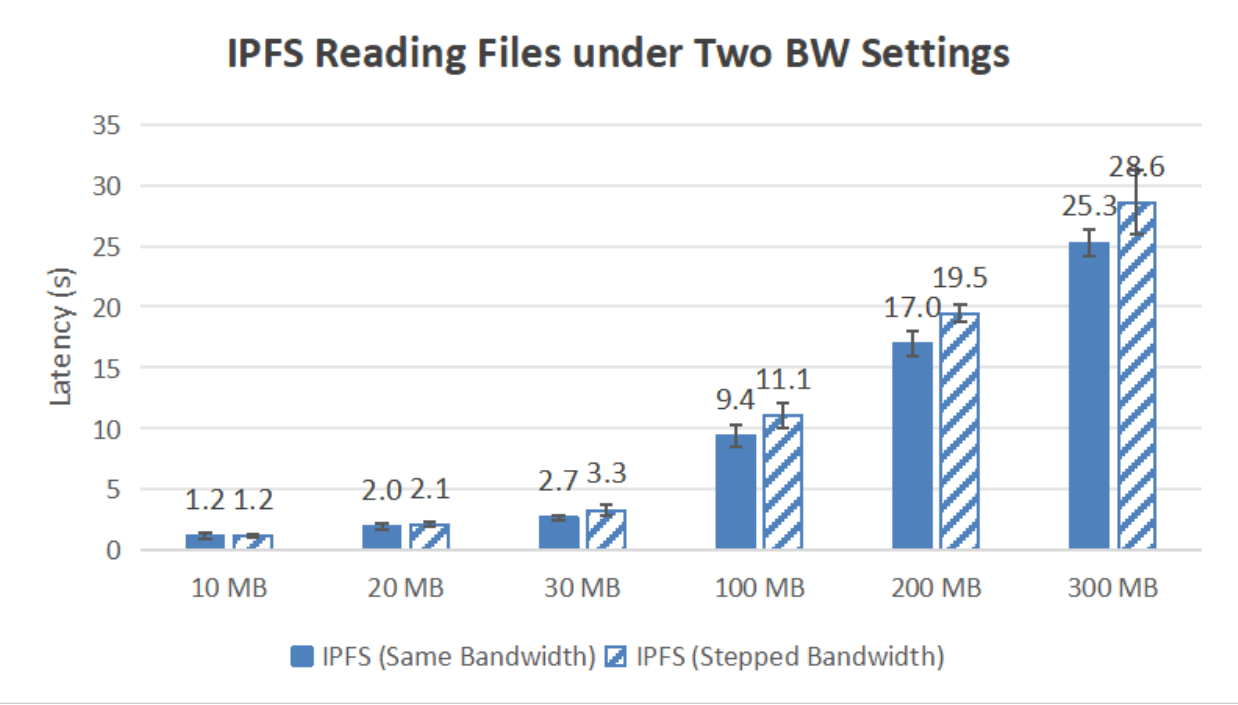}
     \caption {Reading latency of IPFS under different bandwidth settings}
     \label{ipfs_r}
\end{figure}

In the last experiment, we adopted a stepped bandwidth setting to simulate varying network conditions scenarios among different organizations.
The bandwidths of nodes in four organizations are limited to 400, 800, 1,200, and 1,600 Mbps, respectively. Note that the average bandwidth is 1,000 Mbps, the same as in the second experiment setting. We observe the performance of IPFS and the DBNode system by comparing the latency under the stepped bandwidth setting and the same bandwidth setting.

Fig.~\ref{ipfs_w} and \ref{ipfs_r} display the result of IPFS. As shown in Fig.~\ref{ipfs_w}, The writing latency of IPFS remains unchanged under both settings. This is because the writing operation is performed locally in IPFS, and the network condition does not impact the writing performance. However, Fig.~\ref{ipfs_r} shows that the reading latency increases when using the stepped bandwidth setting. 

A significant increase in reading latency is evident, especially with large files. This is because IPFS treats all nodes equally and does not optimize for different nodes. Nodes with lower bandwidth can noticeably slow down the process of fetching files. In particular, obtaining large files is more sensitive to bandwidth, leading to a more pronounced increase in reading latency.


\begin{figure}[!htbp]
    \centering
     \includegraphics[width=10 cm, height=6 cm]{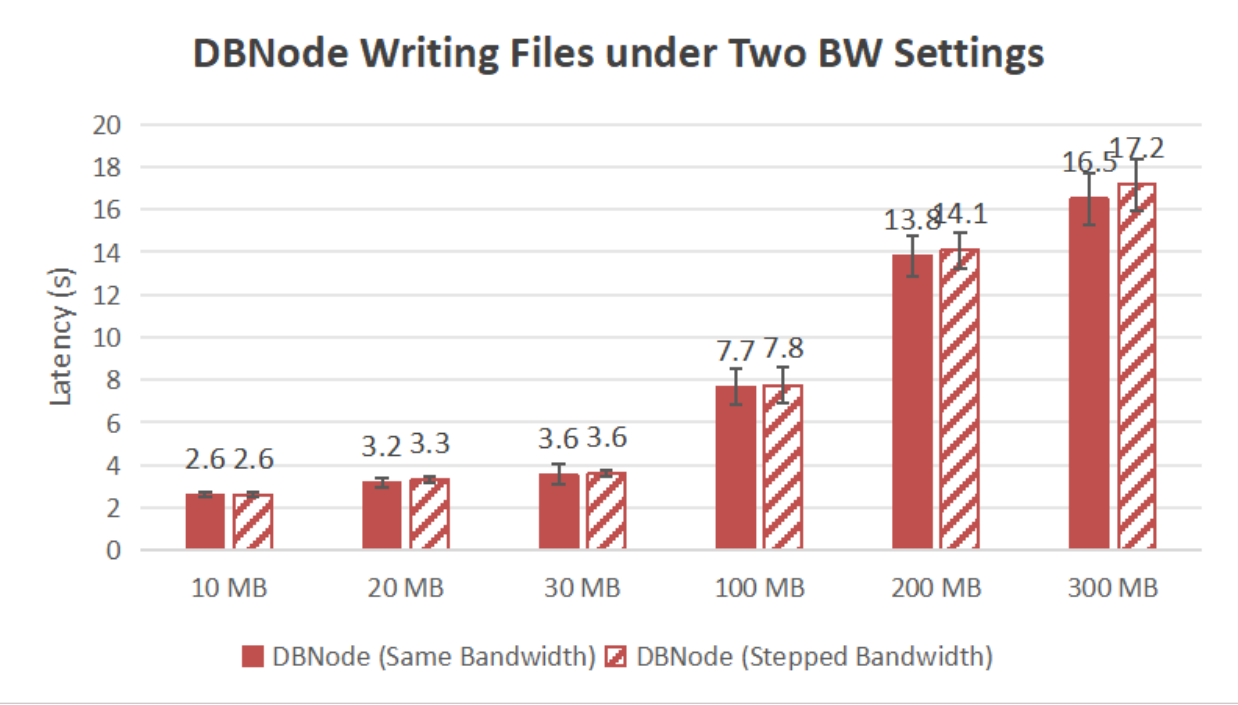}
     \caption {Writing latency of DBNode under different bandwidth settings}
     \label{db_w}
\end{figure}
\begin{figure}[!htbp]
    \centering
     \includegraphics[width=10 cm, height=6 cm]{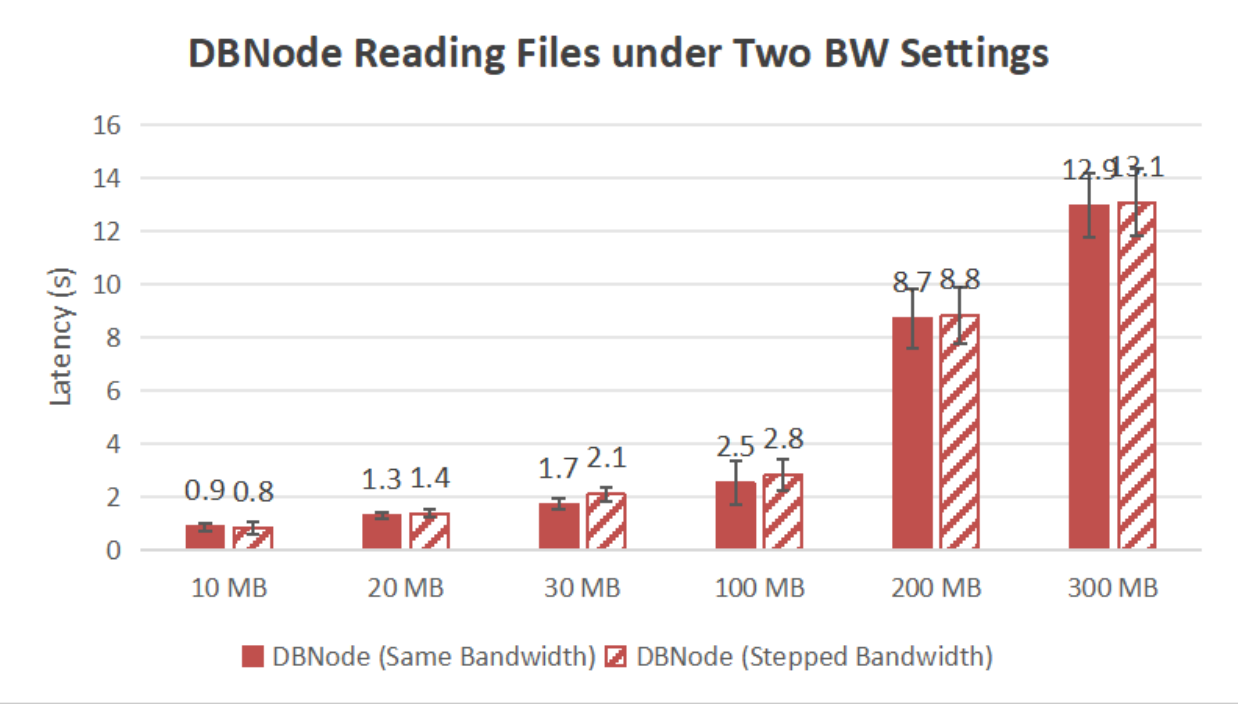}
     \caption {Reading latency of DBNode under different bandwidth settings}
     \label{db_r}
\end{figure}

A slight increase in writing latency is observed in the DBNode system, as shown in Fig.~\ref{db_w}. Different from IPFS, the writing operation in this system involves network communication. In that case, the communication between the client and nodes with low bandwidth becomes the system bottleneck. However, due to the usage of erasure coding and hash slots, files are divided into small-size chunks, and nodes with lower bandwidth are allocated with fewer hash slots. Consequently, only a limited number of small-size chunks are stored on nodes with low bandwidth, mitigating the impact on writing latency.

Fig.~\ref{db_r} demonstrates that the DBNode system effectively adapts to changes in network bandwidth compared to IPFS. The reading latency remains almost the same, although the bandwidth setting is changed while the reading latency of IPFS increases in this scenario. Two key factors may cause such a difference. First, the DBNode system allocates more hash slots to nodes with high bandwidth, resulting in more stored chunks in these nodes and reduced impact from nodes with lower bandwidth. Second, a ($n$, $k$) erasure code allows the client to recover the file based on the first $k$ obtained chunks. In that case, the client will not wait for other chunks that might not have been transferred by the nodes with low bandwidth.


\section{Conclusion}
This research presents an innovative solution for storing large files on the blockchain. Our design is decentralized and aligns with core blockchain principles. By eliminating the need for clients to manage off-chain storage systems and encapsulating the complexities, we make the use of blockchain for large files as seamless as storing regular transactions.
Throughout our experiments, we observed a slight storage overhead and minor latency when writing files. It is essential to emphasize that these obstacles are of negligible consequence in comparison to the substantial advantages offered by DBNode. The modest storage overhead primarily arises from the mirror strategy, which ensures the availability and accessibility of file chunks across the network. 
Despite the minor latency in writing files, our system affords substantial benefits by enhancing privacy and granting clients full control over file access. This benefit is accomplished by implementing smart contracts and incorporating an erasure coding mechanism, allowing clients to determine their preferred file storage nodes and access rules. Since file storage is a one-time event while file retrieval occurs frequently, minimizing latency during file retrieval is crucial. According to our experiments, DBNode outperforms IPFS regarding file retrieval speed and latency.
In future work, our research will focus on conducting comparative examinations with other solutions and improving writing file speed.

\bibliographystyle{ieeetr}
\bibliography{conference}

\hfill



\end{document}